%% file: main.tex
\documentclass[aps,prl,twocolumn,superscriptaddress,floatfix,reprint,preprintnumbers]{revtex4-2}

\usepackage{amsmath,amssymb,amsfonts}
\usepackage{graphicx}
\usepackage{hyperref}
\usepackage{xcolor}
\usepackage{tikz}
\usepackage{subcaption}
\usetikzlibrary{calc,arrows,decorations.markings,shapes.misc,decorations.pathmorphing}
\usepackage{CJK}
\usepackage{makecell}
\usepackage{booktabs,tabularx}

\newcommand{\cN}{\mathcal{N}}
\newcommand{\cF}{\mathcal{F}}
\newcommand{\cR}{\mathcal{R}}

\def\boxsize{0.55}
\def\blobsize{0.45cm}
\def\bwblobsize{0.28cm}
\newcommand{\blackdot}{\node[circle, fill=black, draw, inner sep=0pt, minimum size=\bwblobsize]}
\newcommand{\whitedot}{\node[circle, fill=white, draw, inner sep=0pt, minimum size=\bwblobsize]}
\newcommand{\greyblob}{\node[circle, fill=black!20, draw, inner sep=1pt, minimum size=\blobsize]}
\newcommand{\drawULblack}{\blackdot (UL) at (-\boxsize,  \boxsize) {};}
\newcommand{\drawURblack}{\blackdot (UR) at ( \boxsize,  \boxsize) {};}

\newcommand{\drawLLblack}{\blackdot (LL) at (-\boxsize, -\boxsize) {};}
\newcommand{\drawULwhite}{\whitedot (UL) at (-\boxsize,  \boxsize) {};}
\newcommand{\drawURwhite}{\whitedot (UR) at ( \boxsize,  \boxsize) {};}
\newcommand{\drawLLwhite}{\whitedot (LL) at (-\boxsize, -\boxsize) {};}
\newcommand{\drawUL}[1]{\greyblob (UL) at (-\boxsize,  \boxsize) {#1};}
\newcommand{\drawUR}[1]{\greyblob (UR) at ( \boxsize,  \boxsize) {#1};}
\newcommand{\drawLR}[1]{\greyblob (LR) at ( \boxsize, -\boxsize) {#1};}

\newcommand{\drawboxinternallines}{\draw (UL) -- (UR) -- (LR) -- (LL) -- (UL);}

\begin{document}
\preprint{MPP-2026-88}

\title{Bootstrapping the Four-Point NMHV Stress-Tensor Form Factor}

\begin{CJK*}{UTF8}{}
\CJKfamily{gbsn}
\author{Song He (何颂)}
\email{songhe@itp.ac.cn}
\affiliation{New Cornerstone Laboratory, Institute of Theoretical Physics, Chinese Academy of Sciences, Beijing 100190, China}
\affiliation{School of Fundamental Physics and Mathematical Sciences, Hangzhou Institute for Advanced Study, UCAS \& ICTP-AP, Hangzhou, 310024, China}
\author{Jiahao Liu (刘家昊)}
\email{liujiahao@itp.ac.cn}
\affiliation{New Cornerstone Laboratory, Institute of Theoretical Physics, Chinese Academy of Sciences, Beijing 100190, China}
\affiliation{School of Physical Sciences, University of Chinese Academy of Sciences, No. 19A Yuquan Road, Beijing 100049, China}
\author{Qinglin Yang (杨清霖)}
\email{qlyang@mpp.mpg.de}
\affiliation{Max Planck Institut f\"ur Physik, Werner Heisenberg Institut, D-85748 Garching bei M\"unchen, Germany}

\begin{abstract}
We bootstrap the four-point next-to-maximally helicity-violating (NMHV) form factor of the chiral stress-tensor supermultiplet in planar maximally supersymmetric Yang-Mills theory through three loops at the symbol level. At two loops, an ansatz built from NMHV leading singularities and the relevant five-point one-mass integral function space is fixed uniquely by physical constraints; the resulting ratio function symbol contains 78 letters, all drawn from the 88-letter alphabet previously identified in the four-point MHV sector. At three loops, using this 88-letter alphabet as input and imposing extended Steinmann relations satisfied by the minimally-subtracted hard function, together with other physical constraints, we determine the three-loop symbol uniquely. Both results pass soft, double-soft and directional dual conformal invariance (DDCI) checks, provide the first multi-loop non-MHV form-factor data, and support the universality of the 88-letter alphabet for four-point form factors beyond the MHV sector.
\end{abstract}

\maketitle
\end{CJK*}

\section{Introduction}
Gluon fusion is the dominant production mechanism for Higgs bosons at colliders. Consequently, the corresponding scattering amplitudes are key observables for testing the Standard Model and probing new physics. In the large-top-mass limit, Higgs-plus-parton amplitudes can be reduced to form factors of gauge-invariant effective operators \cite{Wilczek:1977zn,Shifman:1978zn}, making the analytic calculation and investigation of form factors an essential component of particle-physics studies.

In modern scattering-amplitude research, supersymmetric form factors of the chiral stress-tensor supermultiplet \cite{vanNeerven:1985ja} (half-BPS form factors) in planar maximally supersymmetric Yang-Mills theory ($\mathcal{N}{=}4$ sYM) have become a useful counterpart of the Higgs-production form factor; see the recent review \cite{Arkani-Hamed:2022rwr}. They retain enough structure from scattering amplitudes to admit many of the same modern tools, while the off-shell operator insertion introduces genuinely new features. Examples include the form factor/periodic Wilson-loop duality \cite{Maldacena:2010kp,Alday:2007he,Brandhuber:2010ad,Brandhuber:2011tv,Gao:2013dza}, which has been used to compute two-loop MHV form factors at all multiplicities \cite{Li:2024rkq}, recursion relations for tree-level objects and loop integrands~\cite{Bianchi:2018peu}, Grassmannian geometry and on-shell diagrams for leading singularities~\cite{Frassek:2015rka,Bork:2016hst,Bork:2016xfn}, color-kinematics duality \cite{Boels:2012ew,Yang:2016ear,Lin:2021kht,Lin:2021lqo}, the form factor operator product expansion (OPE) \cite{Sever:2020jjx,Sever:2021nsq,Sever:2021xga,Basso:2023gfu}, and the Steinmann/cluster bootstrap program (see the review \cite{Caron-Huot:2020bkp}). For form factors of the stress-tensor supermultiplet, or equivalently the half-BPS $\mathrm{Tr}(\phi^2)$ multiplet, this bootstrap program has enabled the determination of the three-point MHV form factor up to eight loops and the four-point MHV form factor up to four loops \cite{Dixon:2020bbt,Dixon:2022xqh,Dixon:2022rse,Dixon:2024yvq,Guo:2024bsd,MHV:fourloop}; see also \cite{Guan:2023gsz} for the three-loop three-point non-planar correction. The related three-point MHV form factor of the $\mathrm{Tr}(\phi^3)$ operator has been calculated to six loops~\cite{Basso:2024hlx,Henn:2024pki}; see also \cite{Guo:2021bym,Guo:2022pdw} for the two-loop four-point case.

Beyond their calculational accessibility, half-BPS form factors have revealed a variety of hidden symmetries and analytic structures. At tree level and one loop, they have been shown to possess a dual-conformal-invariance structure~\cite{Bianchi:2018rrj}. At higher perturbative orders, the three-point MHV form factor is organized by a $C_2$-type cluster algebra, while the four-point MHV form factor involves, through four loops, a stable 88-letter symbol alphabet. More strikingly, the three-point MHV form factor was found to be antipodally dual to the six-point MHV amplitude \cite{Dixon:2021tdw,Dixon:2023kop}. A closely related antipodal self-duality was later uncovered in the four-point MHV form factor \cite{Dixon:2022xqh}, offering an explanation for the three-point duality. Furthermore, three-point MHV form factors were found to coincide with the maximally transcendental parts of $H\to ggg$ amplitudes \cite{Brandhuber:2012vm,Duhr:2012fh,Chen:2025utl}, providing another important example of the principle of maximal transcendentality between $\mathcal{N}{=}4$ sYM and QCD \cite{Kotikov:2002ab,Kotikov:2004er,Dixon:2017nat}. The study of half-BPS form factors is therefore expected to reveal, in a controlled setting, deeper mathematical and physical structures underlying collider processes involving Higgs production.

These advances have so far been concentrated mainly in the MHV sector. In this work, we extend the study of half-BPS form factors beyond the MHV sector by bootstrapping the four-point NMHV form factor at the symbol level~\cite{Goncharov:2010jf,Duhr:2011zq}. For non-MHV sectors, one encounters nontrivial leading-singularity prefactors and a richer set of physical constraints and limits. We organize the ansatz in terms of three classes of form-factor leading singularities as kinematic prefactors. At two loops, instead of constructing all integrable symbols directly from the 113 letters of the master integrals, we build on the known five-point one-mass function space \cite{Abreu:2020jxa,Chicherin:2021dyp,Abreu:2021smk,Abreu:2023rco} constructed by the canonical differential equation method \cite{Henn:2013pwa,Henn:2014qga}. Physical consistency conditions then fix the two-loop result uniquely; its symbol contains 78 letters, all drawn from the 88-letter alphabet of the four-point MHV form factors up to four loops.

This observation provides the starting point for the three-loop bootstrap. We assume that the same 88-letter alphabet applies to the three-loop minimally-subtracted hard function. We then construct the symbol space from these letters and impose the extended Steinmann relations \cite{Caron-Huot:2019vjl,Caron-Huot:2019bsq,He:2021mme} on the resulting weight-six symbol space. Together with the leading-singularity decomposition and the physical constraints used at lower-loop order, these conditions fix the symbol of the three-loop NMHV hard function.

The resulting two- and three-loop symbols pass several independent consistency checks, including soft and double-soft limits and directional dual conformal invariance. They provide evidence that four-point form factors, in both MHV and NMHV sectors, are governed by the same 88-letter alphabet, with their hard functions obeying analogous extended Steinmann relations. These results provide the first multi-loop non-MHV form-factor data and open a path toward higher-loop four-point form factors and their underlying mathematical structure.

\section{NMHV Form Factor Ratio Function and Leading Singularities}
We consider the four-point super form factor in the large-$N_c$ limit~\cite{Brandhuber:2011tv,Bork:2012tt},
\begin{equation}
    \mathcal{F}_4(p_i,q,\eta_i):=\int{\rm d}^4x {\rm d}^4\theta^+\ \text{e}^{{-}\text{i}(q x{+}\theta^+\gamma)}\langle \Omega_4|\mathcal{T}(x,\theta^+)|0\rangle
\end{equation}
where the chiral stress-tensor multiplet is $\mathcal{T}(x,\theta^+)=\text{Tr}(\phi(x)^2)+\cdots+(\theta^+)^4\mathcal{L}(x)$, with $\mathcal{L}(x)$ denoting the chiral on-shell Lagrangian of $\mathcal{N}=4$ sYM, and $\langle\Omega_4|$ is the four-particle external superstate depending on $\{p_i,\eta_i\}$. This quantity has a natural decomposition \cite{Bork:2011cj,Brandhuber:2011tv}
\begin{equation}
    \mathcal{F}_4=\mathcal{F}_{4,0}+\mathcal{F}_{4,1}+\mathcal{F}_{4,2},
\end{equation}
where $\mathcal{F}_{4,k}$ denotes the \(\mathrm{N}^k\mathrm{MHV}\) sector and has Grassmann degree $4k+8$. The MHV sector $\mathcal{F}_{4,0}$ has been studied previously in the bootstrap literature \cite{Dixon:2022xqh} and has recently been extended to four loops \cite{MHV:fourloop}. We define the ratio function by
\begin{equation}
    \cF_{4,1} = \cF_{4,0} \, \cR_4 \, .
\end{equation}
The two-loop bootstrap below is performed directly for the infrared-finite ratio function $\cR_4^{(2)}$, while at three loops we use the associated minimally-subtracted hard function, obtained by subtracting the universal infrared singularities from the form factor.

The kinematics of $\mathcal{F}_{4,1}$ depends on four massless external legs $p_i$ and one off-shell operator momentum,
\(
q=\sum_{i=1}^4 p_i
\),
with $p_i^2=0$ and $q^2\neq0$.
We use the eight dimensionless invariants~\cite{Dixon:2022xqh}
\begin{equation}
    \begin{aligned}
    u_i =
    \frac{(p_i+p_{i+1})^2}{q^2},\ \ 
    v_i =
    \frac{(p_i+p_{i+1}+p_{i+2})^2}{q^2}\,,
    \end{aligned}
\end{equation}
with $1\leq i\leq4$, which obey three independent linear relations: $-u_1+u_3+v_4+v_1=1$ together with two independent cyclic images; these leave five independent kinematic variables. In practice, we implement the kinematics using periodic momentum twistors and an OPE parametrization \cite{Sever:2020jjx,Sever:2021nsq,Sever:2021xga,Basso:2023gfu}; explicit parametrizations are given in the Supplemental Material.

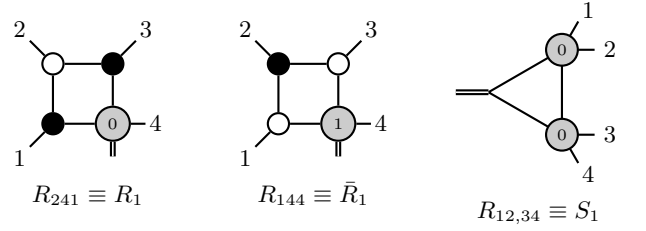
\begin{figure}[t]
    \centering
    \begin{minipage}{0.31\linewidth}
        \centering
        \begin{tikzpicture}[thick, scale=0.72]
          \drawLLblack
          \drawULwhite
          \drawURblack
          \drawLR{$\scriptstyle{0}$}
          \drawboxinternallines
          \draw (UL) -- ++(135:0.6) node[above left=-2pt] {$2$};
          \draw (UR) -- ++(45:0.6) node[above right=-2pt] {$3$};
          \draw (LR) -- ++(0:0.6) node[right=-2pt] {$4$};
          \draw[double] (LR) -- ++(-90:0.6);
          \draw (LL) -- ++(-135:0.6) node[below left=-2pt] {$1$};
        \end{tikzpicture}

        $R_{241}\equiv R_1$
    \end{minipage}\hfill
    \begin{minipage}{0.31\linewidth}
        \centering
        \begin{tikzpicture}[thick, scale=0.72]
          \drawLLwhite
          \drawULblack
          \drawURwhite
          \drawLR{$\scriptstyle{1}$}
          \drawboxinternallines
          \draw (UL) -- ++(135:0.6) node[above left=-2pt] {$2$};
          \draw (UR) -- ++(45:0.6) node[above right=-2pt] {$3$};
          \draw (LR) -- ++(0:0.6) node[right=-2pt] {$4$};
          \draw[double] (LR) -- ++(-90:0.6);
          \draw (LL) -- ++(-135:0.6) node[below left=-2pt] {$1$};
        \end{tikzpicture}

        $R_{144}\equiv \bar{R}_1$
    \end{minipage}\hfill
    \begin{minipage}{0.31\linewidth}
        \centering
        \begin{tikzpicture}[thick, scale=0.72]
        \coordinate (T1) at ( 0.45 ,  0.78);
        \coordinate (T2) at ( 0.45 , -0.78);
        \coordinate (T3) at (-0.9 ,  0.0 );
        \draw (T1) -- (T2) -- (T3) -- (T1);
        \draw (T1) -- ++( 60:0.6) node[above right=-2pt] {$1$};
        \draw (T1) -- ++(  0:0.6) node[right] {$2$};
        \draw (T2) -- ++(  0:0.6) node[right] {$3$};
        \draw (T2) -- ++(-60:0.6) node[below right=-2pt] {$4$};
        \draw[double] (T3) -- ++(180:0.6);
        \node[circle, fill=black!20, draw, minimum size=0.42cm, inner sep=1pt] at ( 0.45 ,  0.78) {$\scriptstyle{0}$};
        \node[circle, fill=black!20, draw, minimum size=0.42cm, inner sep=1pt] at ( 0.45 , -0.78) {$\scriptstyle{0}$};
        \end{tikzpicture}

        $R_{12,34}\equiv S_1$
    \end{minipage}
    \caption{Three types of leading singularities.}
    \label{fig:leading-singularities}
    \vspace{-3ex}
\end{figure}

Supersymmetric form factors at $L$ loops admit a decomposition into products of supersymmetric kinematic prefactors, known as \textit{leading singularities} or \textit{$R$-invariants} \cite{Drummond:2008vq,Cachazo:2008vp,Bullimore:2009cb}, and weight-$2L$ transcendental functions. At one loop, the four-point NMHV form factor involves three types of leading singularities \cite{Bork:2012tt,Bianchi:2018rrj}: two box-type $R$-invariants $R_1 = R_{241}$ and $\bar{R}_1 = R_{144}$, together with the algebraic $R$-invariant $S_1 = R_{12,34}$, arising from three-mass-triangle cuts. They can be computed from supersymmetric on-shell diagrams, as shown in Fig.~\ref{fig:leading-singularities}. In the diagrams, black/white trivalent vertices denote three-point MHV and \(\overline{\mathrm{MHV}}\) superamplitudes respectively. Gray blobs denote higher-point form-factor/amplitude vertices, and the integer inside each blob labels its N$^k$MHV degree \(k\). A double line marks the off-shell momentum carried by the operator insertion. The prefactor $\bar{R}_1$ is drawn in a different but equivalent on-shell representation compared to Ref.~\cite{Bianchi:2018rrj}, making its parity relation to $R_1$ manifest. We provide more details about these leading singularities and their expressions in terms of momentum twistors in the Supplemental Material.

Motivated by the on-shell-diagram construction and the Grassmannian interpretation of supersymmetric on-shell functions \cite{Arkani-Hamed:2012zlh,Frassek:2015rka}, we conjecture that these three classes of leading singularities suffice to organize the four-point NMHV form factor to \textit{all loop orders} \footnote{We note that the leading singularities of the form factor at all loop orders, along with their corresponding on-shell diagrams, can always be obtained from the one-loop on-shell functions through gluing operations, such as BCFW bridges and inverse-soft factors \cite{Arkani-Hamed:2012zlh}. Similar to the case of scattering amplitudes, these operations do not introduce additional $R$-invariants beyond one-loop leading singularities. The successful two- and three-loop bootstraps below provide nontrivial evidence for this conjecture. }.

As a consequence of our conjecture, $\cR_4$ at arbitrary $L$ loops
always admits the decomposition
\begin{equation}\label{eq:allloop}
\begin{aligned}
    \cR_4^{(L)}
    = \sum_{i=1}^4 \Bigl(
        R_i\,f_i^{(L)}
        + \bar{R}_i\,\bar{f}_i^{(L)}
      \Bigr)
      + \sum_{i=1}^2 S_i\,f_{s,i}^{(L)} \, ,
\end{aligned}
\end{equation}
where $R_i$, $\bar{R}_i$ and $S_i$ are cyclic images of $R_1$, $\bar{R}_1$, $S_1$ respectively, and $f_a^{(L)}$ are weight-$2L$ transcendental functions. Since $R_1$ and $\bar{R}_1$ are related by parity, it is natural to introduce the combinations
\begin{equation}
    R^\pm_i=R_i \pm \bar{R}_i,\ \ f_{\pm,i}^{(L)}=\frac{f_{i}^{(L)}\pm\bar{f}_{i}^{(L)}}{2} \, ,\ i=1,\ldots,4.
\end{equation}
In this basis the cyclic ansatz becomes
\begin{equation}\label{eq:cycansatz}
\begin{aligned}
    \cR_4^{(L)}
    = \sum_{i=1}^4
    \Bigl(
        R_{i}^{+} f_{+,i}^{(L)}
        + R_{i}^{-} f_{-,i}^{(L)}
    \Bigr)
    + \sum_{i=1}^2 S_{i} f_{s,i}^{(L)} \, .
\end{aligned}
\end{equation}
With this choice, $f_{+,i}^{(L)}$ and $f_{s,i}^{(L)}$ are spacetime parity even, while $f_{-,i}^{(L)}$ is parity odd. We note that the four $R_{i}^{-}$ are not independent. Comparing the BCFW representation of the tree-level ratio function with its parity-conjugate representation gives the linear relation~\cite{Bork:2014eqa}:
\begin{equation}\label{eq:oddrel}
    R_{1}^{-}-R_{2}^{-}
    +R_{3}^{-}-R_{4}^{-}=0\, .
\end{equation}
Therefore, one should eliminate the redundant odd prefactor. After using Eq.~\eqref{eq:oddrel}, the parity-odd part of the ansatz can be written as
\begin{equation}\label{eq:genuine-odd}
   R_{1}^{-}\bigl(f_{-,1}+f_{-,4}\bigr)
   + R_{2}^{-}\bigl(f_{-,2}-f_{-,4}\bigr)
   + R_{3}^{-}\bigl(f_{-,3}+f_{-,4}\bigr)\,.
\end{equation}
In other words, three independent combinations of the four functions $f_{-,i}$ already capture all physical information of the parity-odd part.

Before turning to the explicit bootstrap, we recall the tree- and one-loop results. At tree level, only the parity-even functions are present (for $i=1,2,3,4$),
\begin{equation}
    f_{+,i}^{(0)}=\frac{1}{2}\,,
    \qquad
    f_{-,i}^{(0)}=f_{s,i}^{(0)}=0\, .
\end{equation}
\textit{i.e.} the tree-level NMHV form factor is given by the cyclic sum of the parity-even leading singularities alone. At one loop, the parity-odd function is still absent, while the parity-even and algebraic sectors are given by weight-two functions~\cite{Bork:2012tt,Bianchi:2018peu},
\begin{equation}
    f_{-,i}^{(1)}=0\,,
    \qquad
    f_{+,i}^{(1)}=V_i\,,
    \qquad
    f_{s,i}^{(1)}=T_i\, ,
\end{equation}
where $V_i$ denotes the corresponding finite part of a one-mass box function and two two-mass-hard box functions~\cite{Bianchi:2018rrj}, while $T_i$ is the corresponding finite three-mass triangle function.

\section{Two-Loop bootstrap}
\begin{figure}[t]
	\centering
	\begin{subfigure}{0.15\textwidth}\centering
		\includegraphics[scale=0.25]{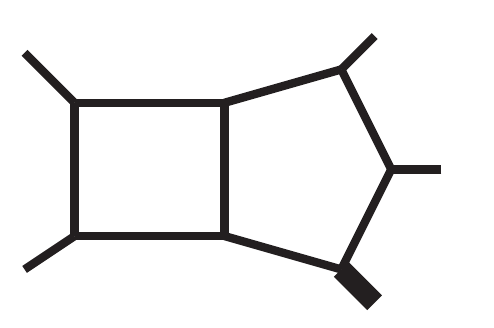}
		\caption{PBmzz} 
		\label{fig:pbmzz}
	\end{subfigure}
	\begin{subfigure}{0.15\textwidth}\centering
		\includegraphics[scale=0.25]{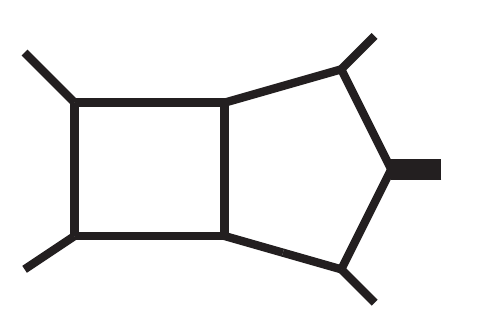}
		\caption{PBzmz}
		\label{fig:pbzmz}
	\end{subfigure}
	\begin{subfigure}{0.15\textwidth}\centering
		\includegraphics[scale=0.25]{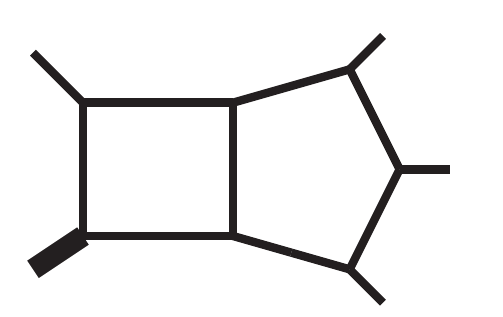}
		\caption{PBzzz}
		\label{fig:pbzzz}
	\end{subfigure}
	\begin{subfigure}{0.15\textwidth}\centering
		\includegraphics[scale=0.25]{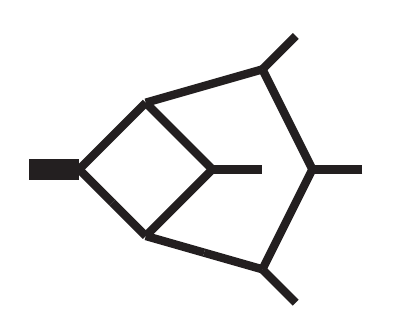}
		\caption{HBzzz}
		\label{fig:hbzzz}
	\end{subfigure}
	\begin{subfigure}{0.20\textwidth}\centering
		\includegraphics[scale=0.25]{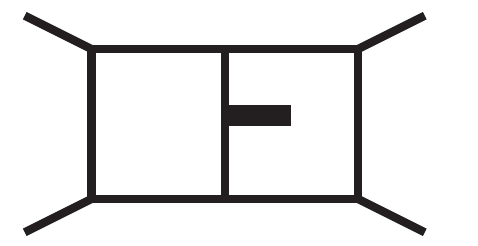}
		\caption{DPzz}
		\label{fig:zz}
	\end{subfigure}
	\caption{Two-loop five-point one-mass Feynman integral families for form factor bootstrap. } 
	\label{fig_families_int}
    \vspace{-2ex}
\end{figure}

\paragraph*{\underline{1. Two-loop function space}} At two loops, rather than constructing symbols from an alphabet, we start from the known symbol space of two-loop five-point one-mass Feynman integrals~\cite{Abreu:2020jxa,Chicherin:2021dyp,Abreu:2021smk,Abreu:2023rco}, supplemented by the relevant products of lower-weight functions. For the non-planar integral topologies, we only need the topologies in which the non-planar leg is massive, so only the five integral families in Fig.~\ref{fig_families_int} enter the calculation. This is the natural input for the transcendental function basis of the present problem, analogous to previous bootstraps for amplitudes and form factors~\cite{Guo:2021bym,Guo:2022qgv,Carrolo:2025pue,Carrolo:2025agz,Carrolo:2026qpu}. Collecting the symbols of all these functions gives a space of size 989 with 113 symbol letters. The relevant algebraic dependence involves four square roots
\begin{equation}
   \left\{ r_1{=}\sqrt{\lambda(u_2,u_4,1)},\  \Sigma_1{=}\sqrt{\lambda(u_1 u_4,u_2 u_3,v_1 v_3)}\right\}
\end{equation}
and their cyclic images $\{r_2, \Sigma_2\}$ with $\lambda(x,y,z):=x^2{+}y^2{+}z^2{-}2xy{-}2xz{-}2yz$, together with the parity-odd Lorentz invariant
$\text{tr}_5:=4i\epsilon_{\mu\nu\rho\sigma}p_1^\mu p_2^\nu p_3^\rho p_4^\sigma$. Since we are bootstrapping the ratio function, however, the final answer must be a finite combination of master integrals with all infrared divergences canceled, which reduces the initial space to 638. Combined with the cyclic ansatz, this gives an initial count of $3\times 638=1914$ unknowns.

\paragraph*{\underline{2. Fixing the ansatz}} We now constrain this ansatz by imposing symmetry, spurious-pole cancellation, and multi-collinear consistency conditions. 

We first impose the symmetry constraints, consisting of Galois invariance under sign flips of the algebraic roots, spacetime parity, and dihedral symmetry. We then impose spurious-pole cancellation. Up to dihedral symmetry, the leading singularities have one type of unphysical pole, $\langle 3,4,1^+,3^+\rangle\to0$, written in the periodic momentum-twistor notation reviewed in the Supplemental Material. Cancellation of this unphysical pole requires the full ratio function to satisfy
\begin{equation}
    f_{4}-\bar{f}_1+f_{s,1}\biggr|_{\langle 3,4,1^+,3^+\rangle\to0}\to0 \, .
\end{equation}

\begin{table}[t]
    \caption{Numbers of unknowns for two-loop ratio function bootstrap.}
    \label{tab:bootstrap}
    \centering
    \small
    \begin{ruledtabular}
    \begin{tabular}{lc}
        Condition &  Unknowns left \\
        \hline
        \addlinespace[1pt]
        Finiteness & 1914 \\
        \addlinespace[1pt]
        Galois and parity symmetry &  605 \\
        \addlinespace[1pt]
        Dihedral symmetry &  308 \\
        \addlinespace[1pt]
        Spurious-pole cancellation &  88 \\
        \addlinespace[1pt]
        Collinear limit  &  7 \\
        \addlinespace[1pt]
        Triple-collinear limit &  0 \\
        \addlinespace[1pt]
        Soft limit &  0 \\
        \addlinespace[1pt]
        Double-soft limit &  0 \\
        \addlinespace[1pt]
          Directional dual conformal invariance & 0\\ \addlinespace[0pt]
	    \end{tabular}
	    \end{ruledtabular}
        \vspace{-3ex}
	\end{table}

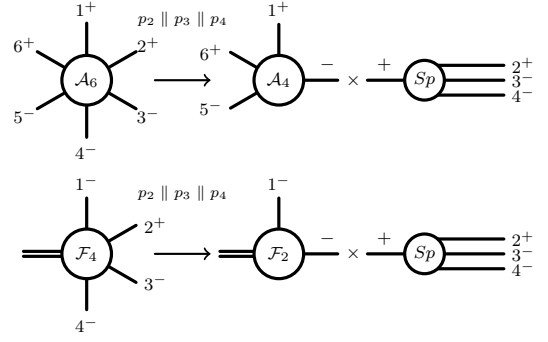
\begin{figure}[!tbp]
    \centering
    \begin{tikzpicture}[
        line width=0.6pt,
        scale=0.82,
        transform shape,
        line cap=round,
        line join=round,
        every node/.style={font=\footnotesize},
        object/.style={circle,draw,very thick,fill=white,minimum size=0.80cm,inner sep=0pt},
        split/.style={circle,draw,very thick,fill=white,minimum size=0.64cm,inner sep=0pt}
    ]
        \coordinate (A) at (1.0,1.35);
        \foreach \ang/\lab in {90/1^+,30/2^+,-30/3^-,-90/4^-,-150/5^-,150/6^+}{
            \draw[very thick] ($(A)+(\ang:0.40)$) -- ($(A)+(\ang:0.92)$);
            \node at ($(A)+(\ang:1.15)$) {$\lab$};
        }
        \node[object] at (A) {$\mathcal{A}_6$};

        \draw[->,thick] (2.10,1.35) -- (3.00,1.35)
            node[midway,yshift=28pt,fill=white,inner sep=1.5pt,font=\scriptsize]
            {$p_2\parallel p_3\parallel p_4$};

        \coordinate (aL) at (4.10,1.35);
        \draw[very thick] ($(aL)+(90:0.40)$) -- ($(aL)+(90:0.90)$) node[above] {$1^+$};
        \draw[very thick] ($(aL)+(150:0.40)$) -- ($(aL)+(150:0.90)$) node[left] {$6^+$};
        \draw[very thick] ($(aL)+(-150:0.40)$) -- ($(aL)+(-150:0.90)$) node[left] {$5^-$};
        \draw[very thick] ($(aL)+(0:0.40)$) -- ($(aL)+(0:0.95)$)
            node[pos=0.70,yshift=7pt] {$-$};
        \node[object] at (aL) {$\mathcal{A}_4$};

        \node at (5.30,1.35) {$\times$};

        \coordinate (aR) at (6.45,1.35);
        \draw[very thick] ($(aR)+(180:0.32)$) -- ($(aR)+(180:0.92)$)
            node[pos=0.55,yshift=7pt] {$+$};
        \draw[very thick] ($(aR)+(0,0.24)$) -- ++(1.28,0) node[right] {$2^+$};
        \draw[very thick] (aR) -- ++(1.28,0) node[right] {$3^-$};
        \draw[very thick] ($(aR)+(0,-0.24)$) -- ++(1.28,0) node[right] {$4^-$};
        \node[split] at (aR) {$Sp$};

        \coordinate (F) at (1.0,-1.45);
        \draw[very thick] ($(F)+(-1.02,0.04)$) -- ($(F)+(-0.40,0.04)$);
        \draw[very thick] ($(F)+(-1.02,-0.04)$) -- ($(F)+(-0.40,-0.04)$);
        \draw[very thick] ($(F)+(90:0.40)$) -- ($(F)+(90:0.90)$) node[above] {$1^-$};
        \draw[very thick] ($(F)+(30:0.40)$) -- ($(F)+(30:0.92)$) node[right] {$2^+$};
        \draw[very thick] ($(F)+(-30:0.40)$) -- ($(F)+(-30:0.92)$) node[right] {$3^-$};
        \draw[very thick] ($(F)+(-90:0.40)$) -- ($(F)+(-90:0.90)$) node[below] {$4^-$};
        \node[object] at (F) {$\mathcal{F}_4$};

        \draw[->,thick] (2.10,-1.45) -- (3.00,-1.45)
            node[midway,yshift=28pt,fill=white,inner sep=1.5pt,font=\scriptsize]
            {$p_2\parallel p_3\parallel p_4$};

        \coordinate (fL) at (4.10,-1.45);
        \draw[very thick] ($(fL)+(-0.95,0.04)$) -- ($(fL)+(-0.40,0.04)$);
        \draw[very thick] ($(fL)+(-0.95,-0.04)$) -- ($(fL)+(-0.40,-0.04)$);
        \draw[very thick] ($(fL)+(90:0.40)$) -- ($(fL)+(90:0.90)$) node[above] {$1^-$};
        \draw[very thick] ($(fL)+(0:0.40)$) -- ($(fL)+(0:0.95)$)
            node[pos=0.70,yshift=7pt] {$-$};
        \node[object] at (fL) {$\mathcal{F}_2$};

        \node at (5.30,-1.45) {$\times$};

        \coordinate (fR) at (6.45,-1.45);
        \draw[very thick] ($(fR)+(180:0.32)$) -- ($(fR)+(180:0.92)$)
            node[pos=0.55,yshift=7pt] {$+$};
        \draw[very thick] ($(fR)+(0,0.24)$) -- ++(1.28,0) node[right] {$2^+$};
        \draw[very thick] (fR) -- ++(1.28,0) node[right] {$3^-$};
        \draw[very thick] ($(fR)+(0,-0.24)$) -- ++(1.28,0) node[right] {$4^-$};
        \node[split] at (fR) {$Sp$};
    \end{tikzpicture}
    \caption{Triple-collinear matching between the six-point NMHV amplitude and the four-point NMHV form factor on one super-component.}
    \label{fig:triple-collinear}
    \vspace{-3ex}
\end{figure}

We then impose the multi-collinear limits. In the collinear limit $p_4 \parallel p_3$, the four-point ratio function reduces to the three-point result, which vanishes at any loop order. Since only the prefactors $\bar{R}_{2}$, $\bar{R}_{3}$, and $S_1$ survive in this limit, and all three reduce to the same three-point NMHV tree prefactor, the vanishing of the three-point ratio function requires
\begin{equation}
    \bar{f}_{2}+\bar{f}_{3}+f_{s,1}\biggr|_{p_4 \parallel p_3}\to0 \, .
\end{equation}
We next impose the triple-collinear consistency condition \cite{Catani:1998nv,Catani:2003vu,Badger:2015cxa}: in the limit $p_4 \parallel p_3 \parallel p_2$, the four-point NMHV form factor factorizes into the product of a two-point MHV form factor and the $\mathcal{N}{=}4$ triple-collinear splitting function of Grassmann degree four.
The splitting function can be extracted from the triple-collinear limit of six-point NMHV amplitudes \cite{Kosower:2010yk}. The corresponding amplitude/form-factor comparison is illustrated in Fig.~\ref{fig:triple-collinear}.
In this limit, the form factor leading singularities reduce to linear combinations of the six-point NMHV amplitude $R$-invariants $(i)$, $i=1,\ldots,6$, as
\begin{equation}
\begin{aligned}
R^\pm_1&\to (5)\pm (2), & R^{\pm}_2&\to (6) \pm (3), \\
R^\pm_3&\to (1) \pm  (4), & R^\pm_4&\to 0, \\
S_1&\to (1)+(4), & S_2&\to (2)+(5) .
\end{aligned}
\end{equation}
Using the identity $(1)+(3)+(5)=(2)+(4)+(6)$ to eliminate $(6)$, we have the following constraints on the ansatz:
\begin{equation}
\begin{aligned}
& f_{s,1}+f_2 + f_3 \to g_1, & f_{s,2}+\bar{f}_1-f_2\to g_2, \\
&\bar{f}_2 + f_2 \to g_3, & f_{s,1}-f_2 + f_3 \to g_4, \\
& f_{s,2}+f_1+f_2\to g_5. 
\end{aligned}
\end{equation}
Here $g_i$ denotes the transcendental function multiplying the six-point NMHV amplitude invariant $(i)$ in the known two-loop data \cite{Kosower:2010yk}. Together with the preceding constraints, the ansatz is determined uniquely; the dimension counts at each step are summarized in Table~\ref{tab:bootstrap}.

\paragraph*{\underline{3. Two-loop result}} The resulting two-loop ratio function passes several independent checks. It vanishes in the soft limit, and in the double-soft limit it reduces to the $(+-)$-helicity double soft-gluon current function \cite{Zhu:2020ftr,Czakon:2022dwk} in $\mathcal{N}{=}4$ sYM, which can also be extracted from six-point NMHV amplitudes. Moreover, it is directional-dual-conformally invariant (DDCI) \cite{Bern:2018oao,Chicherin:2018wes} in the \(q^2\to0\) limit, as observed previously in three- and four-point MHV form factors \cite{Lin:2021lqo,Guo:2022qgv,Guo:2024bsd}.

A notable feature of the result is that, although the initial two-loop function space involves 113 symbol letters, the final ratio function symbol contains only 78 letters, all lying inside the 88-letter alphabet observed in the four-point MHV form factor through four loops. Further details of the alphabet are given in the Supplemental Material.

\section{Three-Loop Steinmann Bootstrap}

The containment of the two-loop NMHV symbol letters in the 88-letter MHV alphabet motivates the conjecture that the same alphabet is sufficient for four-point NMHV form factors at {\it all loops}. We test this conjecture by bootstrapping the three-loop hard-function symbol from this 88-letter alphabet.

For this purpose, we use the minimally-subtracted hard function, which is finite after universal infrared subtraction and is related to the ratio function by \(H_{4,1}=H_{4,0}\mathcal R_4\). We construct the weight-six integrable symbol space from the 88-letter alphabet and impose extended Steinmann conditions on this space. Together with the leading-singularity organization and the physical constraints used at lower loop order, these conditions determine the three-loop hard-function symbol uniquely; the dimension counts are summarized in Table~\ref{tab:bootstrap3}. More details are given in the Supplemental Material.

The resulting hard function has the expected soft and double-soft behavior. After converting it back to the ratio function, we also verify DDCI in the \(q^2\to0\) limit. These independent checks support both the leading-singularity organization and the 88-letter alphabet beyond two loops.

\begin{table}[t]
    \caption{Numbers of unknowns for three-loop hard function bootstrap.}
    \label{tab:bootstrap3}
    \centering
    \small
    \begin{ruledtabular}
    \begin{tabular}{lc}
        Condition & Unknowns left \\
        \hline \addlinespace[1pt]
        ES-satisfied integrable symbols & \(54603\times 3\) \\ \addlinespace[1pt]
        Galois, parity, and dihedral symmetry & \(22241\) \\ \addlinespace[3pt]
        \makecell[l]{Spurious-pole cancellation, well-defined \\ multi-collinear limits} & 318 \\ \addlinespace[3pt]
        Eliminate redundancy of the ansatz & 60 \\ \addlinespace[1pt]
        Explicit collinear limit & 21 \\ \addlinespace[1pt]
        Explicit triple-collinear limit & 0 \\ \addlinespace[1pt]
        Explicit soft and double-soft limit & 0 \\ \addlinespace[1pt]
        Directional dual conformal invariance & 0 \\ \addlinespace[1pt]
	    \end{tabular}
	    \end{ruledtabular}
        \vspace{-3ex}
	\end{table}

\section{Discussion and Outlook}
We have bootstrapped the planar four-point NMHV form factor of the chiral stress-tensor supermultiplet in $\cN{=}4$ sYM through three loops at the symbol level.
The two- and three-loop ratio-function and hard-function data are provided in the ancillary files, whose structure is described in the Supplemental Material.

Despite the nontrivial bootstrap procedures through three loops, no letters beyond the 88-letter alphabet of the four-point MHV sector are required. This provides strong evidence for the stability of the 88-letter alphabet for four-point form factors to all loop orders in ${\cal N}{=}4$ sYM, analogous to the roles of $A_3$ and $E_6$ for six- and seven-point amplitudes and $C_2$ for three-point form factors.  One may also ask whether an antipodal duality analogous to the MHV case survives here, which involves only the parity-even part; it would be interesting to understand what object, if any, should be dual to it.

It would also be useful to further study the potential physical implications of our result for Higgs physics, especially Higgs-to-jet amplitudes in the large-top-mass limit. In particular, it would be interesting to obtain the full function-level result and study its relation to the maximal-weight part of recent two-loop Higgs-plus-two-jets results~\cite{DeLaurentis:2026brm}, in analogy with the generalized maximal-weight principle found in \cite{Carrolo:2026qpu}. Several kinematic limits are also worth investigating, including Regge limits \cite{DelDuca:2022skz}, multi-particle factorization \cite{Dixon:2014iba}, light-like limit~\cite{Guo:2022qgv}, and self-crossing kinematics \cite{Dixon:2016epj}, and are expected to yield valuable physical insights into topics such as double parton scattering and BFKL evolution with Higgs production.

\section{Acknowledgments}
We thank Lance Dixon, Johannes Henn and Zhenjie Li for inspiring discussions, and Lance Dixon, Johannes Henn and Gang Yang for helpful comments on the draft. S.H. is supported by the National Natural Science Foundation of China under Grant No. 12225510, 12447101, and by the New Cornerstone Science Foundation.  Q.Y. is supported by the European Union (ERC, UNIVERSE PLUS, 101118787). Views and opinions expressed are however those of the authors only and do not necessarily reflect those of the European Union or the European Research Council Executive Agency. Neither the European Union nor the granting authority can be held responsible for them. J.L. thanks the Max Planck Institute for Physics for its hospitality during the course of this work.

\bibliographystyle{apsrev4-2}
\bibliography{refs}

\input{supplement}

\end{document}

%% file: supplement.tex
\clearpage
\onecolumngrid

\setcounter{section}{0}
\setcounter{equation}{0}
\setcounter{figure}{0}
\setcounter{table}{0}
\renewcommand{\thesection}{S\arabic{section}}
\renewcommand{\theequation}{S\arabic{equation}}
\renewcommand{\thefigure}{S\arabic{figure}}
\renewcommand{\thetable}{S\arabic{table}}
\renewcommand{\theHsection}{supp.\arabic{section}}
\renewcommand{\theHequation}{supp.\arabic{equation}}
\renewcommand{\theHfigure}{supp.\arabic{figure}}
\renewcommand{\theHtable}{supp.\arabic{table}}

\begin{center}
    {\bf Supplemental Material}
\end{center}

\section{Periodic Momentum Twistors and OPE Parametrizations}

Momentum twistors~\cite{Hodges:2009hk} are natural variables for lightlike
dual contours (see also~\cite{ArkaniHamed:2010gh}): a cusp $x_i$ is
represented by the line $(Z_{i-1},Z_i)$, while dual-conformal quantities can be
written in terms of homogeneous four-brackets.  For form factors, the nonzero
operator momentum $q$ makes the dual lightlike contour open after one pass
through the external particles; the contour is instead arranged periodically,
with neighboring periods shifted by $q$.  This is the
periodic Wilson-loop picture
of form factor kinematics
\cite{Maldacena:2010kp,Brandhuber:2010ad,Gao:2013dza}, which we now adapt to
the four-point case considered in the main text.  For the four-point case, we
use dual coordinates
\begin{equation}
    p_i=x_{i+1}-x_i\,,
    \qquad
    q=\sum_{i=1}^4 p_i\,,
    \qquad
    i=1,\ldots,4 .
    \label{eq:supp-dual-x}
\end{equation}
Since $q\neq0$, these dual points do not form a closed polygon.  Following the
periodic prescription for form factors, we have $x_{i+4}=x_i+q$.
Shifted labels such as $i+4$ therefore denote image points rather than labels
reduced modulo four; for the corresponding spinors we take
$\lambda_{i+4}=\lambda_i$.

The momentum twistors are
\begin{equation}
    Z_i=(\lambda_i^\alpha,\mu_i^{\dot\alpha})\,,
    \qquad
    \mu_i^{\dot\alpha}
    =
    x_i^{\alpha\dot\alpha}\lambda_{i\alpha}
    =
    x_{i+1}^{\alpha\dot\alpha}\lambda_{i\alpha}\,.
    \label{eq:supp-Z-def}
\end{equation}
For compactness we denote the images under one positive or negative period by
\begin{equation}
    Z_i^+\equiv Z_{i+4}\,,
    \qquad
    Z_i^-\equiv Z_{i-4}\,.
    \label{eq:supp-Z-images}
\end{equation}
Thus $Z_i^+$ has the same spinor part as $Z_i$ but $\mu_i$ shifted by
$q\lambda_i$.  We write four-brackets as
$\langle i j k l\rangle=\det(Z_i,Z_j,Z_k,Z_l)$.  We also
introduce the infinity twistor, which projects a pair of momentum twistors onto
the ordinary spinor bracket.  With $Z^{\hat A}=(\lambda^\alpha,\mu^{\dot\alpha})$,
we use
\begin{equation}
    (I_\infty)_{\hat A\hat B}
    =
    \begin{pmatrix}
        \epsilon_{\alpha\beta} & 0 \\
        0 & 0
    \end{pmatrix},
    \qquad
    \langle i\,j\rangle
    =
    (I_\infty)_{\hat A\hat B}Z_i^{\hat A}Z_j^{\hat B}
    =
    \epsilon_{\alpha\beta}\lambda_i^\alpha\lambda_j^\beta\, .
    \label{eq:supp-infinity-twistor}
\end{equation}
The standard relation between dual distances and four-brackets is
$x_{ij}^2=\langle i{-}1\,i\,j{-}1\,j\rangle/
(\langle i{-}1\,i\rangle\langle j{-}1\,j\rangle)$, where
$\langle i\,j\rangle$ is the usual spinor bracket.  The variables used in the
main text can therefore be represented as
$u_i=x_{i,i+2}^2/x_{i,i+4}^2$ and
$v_i=x_{i,i+3}^2/x_{i,i+4}^2$, or equivalently
$u_i=\langle i{-}1\,i\,i{+}1\,i{+}2\rangle\langle i{+}3\,i{+}4\rangle/
(\langle i{-}1\,i\,i{+}3\,i{+}4\rangle\langle i{+}1\,i{+}2\rangle)$ and
$v_i=\langle i{-}1\,i\,i{+}2\,i{+}3\rangle\langle i{+}3\,i{+}4\rangle/
(\langle i{-}1\,i\,i{+}3\,i{+}4\rangle\langle i{+}2\,i{+}3\rangle)$.
Here and below, the shifted labels are interpreted using the periodic contour.

We also use an OPE parametrization of the periodic momentum twistors.  The
variables $T_a,S_a,F_a$ are those of the Wilson-loop OPE~\cite{Alday:2010ku}
and its form factor extensions~\cite{Sever:2020jjx,Sever:2021nsq,Sever:2021xga,Basso:2023gfu};
we follow the explicit four-point parametrization of Ref.~\cite{Li:2024rkq}.
The transition between neighboring periods is
\begin{equation}
    Z_{i+4}=\mathsf{P} Z_i\,,
    \label{eq:supp-ope-period}
\end{equation}
with
\begin{equation}
\mathsf{P}=
\begin{pmatrix}
 2 & 1 & 0 & 0 \\
 -1 & 0 & 0 & 0 \\
 0 & 0 & 2 & 1 \\
 0 & 0 & -1 & 0
\end{pmatrix}.
    \label{eq:supp-ope-P}
\end{equation}
The twistors in one period are parametrized as
\begin{equation}
\begin{aligned}
    Z_1&=M_1(0,1,0,1)^{\mathsf T}\,,
    &
    Z_2&=M_1M_2(1,3,-1,1)^{\mathsf T}\,,
    \\
    Z_3&=M_1M_2(1,1,-2,0)^{\mathsf T}\,,
    &
    Z_4&=(0,0,1,0)^{\mathsf T}\,,
\end{aligned}
    \label{eq:supp-ope-Z}
\end{equation}
where
\begin{equation}
M_1=
\begin{pmatrix}
 T_1 & 0 & 0 & 0 \\
 0 & T_1^{-1} & 0 & 0 \\
 0 & 0 & S_1 & 0 \\
 0 & 0 & 0 & S_1^{-1}
\end{pmatrix},
    \label{eq:supp-M1}
\end{equation}
and
\begin{equation}
M_2=
\frac{1}{\sqrt{F_2}S_2T_2}
\begin{pmatrix}
 F_2S_2 & 0 & 0 & 0 \\
 -F_2S_2(T_2^2-1) & F_2S_2T_2^2 & 0 & S_2T_2(S_2-F_2T_2) \\
 T_2-F_2S_2 & 0 & T_2 & 0 \\
 0 & 0 & 0 & S_2^2T_2
\end{pmatrix}.
    \label{eq:supp-M2}
\end{equation}
In this explicit OPE frame, following Ref.~\cite{Li:2024rkq}, the infinity
twistor is represented as the bi-twistor
\begin{equation}
    I_\infty^{\rm OPE}
    =
    \begin{pmatrix}
        1 & -1 & 0 & 0 \\
        0 & 0 & 1 & -1
    \end{pmatrix}^{\mathsf T},
    \qquad
    \langle i\,j\rangle
    =
    \langle i\,j\,I_\infty^{\rm OPE}\rangle\, .
    \label{eq:supp-ope-infinity-twistor}
\end{equation}
Here $\langle i\,j\,I_\infty^{\rm OPE}\rangle$ denotes the four-bracket formed
from $Z_i$, $Z_j$ and the two columns of $I_\infty^{\rm OPE}$.  This choice is
adapted to the transition matrix because
$\mathsf P I_\infty^{\rm OPE}=I_\infty^{\rm OPE}$.
Any dual conformal ratio can be obtained from
the four-brackets of the twistors in Eqs.~\eqref{eq:supp-ope-period}--\eqref{eq:supp-M2},
while spinor brackets are evaluated using Eq.~\eqref{eq:supp-ope-infinity-twistor}.

\section{Form Factor R-Invariants}

Here we collect the form factor leading singularities that enter the ansatz
\eqref{eq:cycansatz}.  We use the standard on-shell superspace variables
$\eta_i^A$, $A=1,\ldots,4$~\cite{Nair:1988bq}.  For form factors of
the stress-tensor multiplet, one also introduces the Grassmann momentum
$\gamma^{A\alpha}$ carried by the operator insertion~\cite{Brandhuber:2011tv}.
The dual Grassmann coordinates are defined by
\begin{equation}
    \theta_{i+1}^{A\alpha}-\theta_i^{A\alpha}
    =
    \lambda_i^\alpha\eta_i^A\,,
    \qquad
    \gamma^{A\alpha}
    =
    \sum_{i=1}^4\lambda_i^\alpha\eta_i^A\, .
    \label{eq:supp-theta-def}
\end{equation}
In parallel, they obey
$\theta_{i+4}^{A\alpha}=\theta_i^{A\alpha}+\gamma^{A\alpha}$.  The corresponding
supertwistors are
\begin{equation}
    \mathcal{Z}_i=(Z_i,\chi_i)\,,
    \qquad
    \chi_i^A=\theta_i^{A\alpha}\lambda_{i\alpha}\,,
    \qquad
    \chi_i^\pm\equiv\chi_{i\pm4}\, .
    \label{eq:supp-supertwistor}
\end{equation}
The basic NMHV building block is the five-bracket
\begin{equation}
    [a,b,c,d,e]
    =
    \frac{
    \delta^{(4)}
    \bigl(
      \chi_a\langle b c d e\rangle
      +\chi_b\langle c d e a\rangle
      +\chi_c\langle d e a b\rangle
      +\chi_d\langle e a b c\rangle
      +\chi_e\langle a b c d\rangle
    \bigr)}
    {\langle a b c d\rangle
     \langle b c d e\rangle
     \langle c d e a\rangle
     \langle d e a b\rangle
     \langle e a b c\rangle}\, .
    \label{eq:supp-five-bracket}
\end{equation}
Here the Grassmann delta function is understood as
\begin{equation}
    \delta^{(4)}(\Xi)=\Xi^1\Xi^2\Xi^3\Xi^4\,,
    \qquad
    \Xi^A =
      \chi_a^A\langle b c d e\rangle
      +\chi_b^A\langle c d e a\rangle
      +\chi_c^A\langle d e a b\rangle
      +\chi_d^A\langle e a b c\rangle
      +\chi_e^A\langle a b c d\rangle\, .
    \label{eq:supp-grassmann-delta}
\end{equation}
Since each $\chi_i^A$ is linear in the on-shell variables $\eta_j^A$, this
numerator is a homogeneous degree-four polynomial in the $\eta$'s, as
appropriate for an NMHV invariant. For the six-point NMHV amplitude, we denote $(1):=[23456]$ and so on, where the six $R$-invariants satisfy $(1)+(3)+(5)=(2)+(4)+(6)$.  

For form factors, the five-brackets involve periodic supertwistors introduced in the previous section. Shifted labels are written as
$i^\pm$ as in Eq.~\eqref{eq:supp-Z-images}. There are two generic on-shell configurations,
\begin{equation}
    \begin{aligned}
    R'_{rst}
    &=
    \vcenter{\hbox{
    \begin{tikzpicture}[thick, scale=0.68, every node/.style={font=\scriptsize}]
      \def\blobsize{0.34cm}
      \drawLLwhite
      \drawUL{$\scriptstyle{0}$}
      \drawUR{$\scriptstyle{0}$}
      \drawLR{$\scriptstyle{0}$}
      \drawboxinternallines
      \draw (UL) -- ++(180:0.65) node[anchor=east] {$r{+}1$};
      \draw[dotted] (UL)+(170:0.5) to [bend left=45] ++(100:0.5);
      \draw (UL) -- ++(90:0.65) node[anchor=south] {$s{-}1$};
      \draw[double] (UR) -- ++(0:0.65);
      \draw (UR) -- ++(45:0.65) node[anchor=south west] {$t{-}1$};
      \draw[dotted] (UR)+(80:0.5) to [bend left=45] ++(50:0.5);
      \draw (UR) -- ++(90:0.65) node[anchor=south] {$s$};
      \draw (LR) -- ++(0:0.65) node[anchor=west] {$t$};
      \draw[dotted] (LR)+(-10:0.5) to [bend left=45] ++(-80:0.5);
      \draw (LR) -- ++(-90:0.65) node[anchor=north] {$r{-}1$};
      \draw (LL) -- ++(-135:0.65) node[anchor=north east] {$r$};
    \end{tikzpicture}}}
    \,,
    R''_{rst}
    &=
    \vcenter{\hbox{
    \begin{tikzpicture}[thick, scale=0.68, every node/.style={font=\scriptsize}]
      \def\blobsize{0.34cm}
      \drawLLwhite
      \drawUL{$\scriptstyle{0}$}
      \drawUR{$\scriptstyle{0}$}
      \drawLR{$\scriptstyle{0}$}
      \drawboxinternallines
      \draw (UL) -- ++(180:0.65) node[anchor=east] {$r{+}1$};
      \draw[dotted] (UL)+(170:0.5) to [bend left=45] ++(100:0.5);
      \draw (UL) -- ++(90:0.65) node[anchor=south] {$s{-}1$};
      \draw (UR) -- ++(0:0.65) node[anchor=west] {$t{-}1$};
      \draw[dotted] (UR)+(80:0.5) to [bend left=45] ++(10:0.5);
      \draw (UR) -- ++(90:0.65) node[anchor=south] {$s$};
      \draw (LR) -- ++(0:0.65) node[anchor=west] {$t$};
      \draw[dotted] (LR)+(-10:0.5) to [bend left=45] ++(-40:0.5);
      \draw (LR) -- ++(-45:0.65) node[anchor=north west] {$r{-}1$};
      \draw[double] (LR) -- ++(-90:0.65);
      \draw (LL) -- ++(-135:0.65) node[anchor=north east] {$r$};
    \end{tikzpicture}}}\, .
    \end{aligned}
    \label{eq:supp-Rprime-Rdoubleprime}
\end{equation}
For $s\neq t$, these two placements of the operator insertion give the same
periodic momentum-twistor invariant after the corresponding region-variable
assignment.  In the notation used in the main text, we suppress the primes and
write
\begin{equation}
    R_{rst}\equiv R'_{rst}=R''_{rst}
    =
    [s{-}1,s,t{-}1,t,r]\, .
    \label{eq:supp-Rrst-general}
\end{equation}
The boundary case of the first configuration is also needed for the four-point
NMHV form factor.  Here the operator sits at a three-point corner, and the
corresponding invariant is
\begin{equation}
    \begin{aligned}
    R'_{rss}
    &=
    \vcenter{\hbox{
    \begin{tikzpicture}[thick, scale=0.68, every node/.style={font=\scriptsize}]
      \def\blobsize{0.34cm}
      \drawLLwhite
      \drawUL{$\scriptstyle{0}$}
      \coordinate (UR) at ( \boxsize,  \boxsize);
      \drawLR{$\scriptstyle{0}$}
      \drawboxinternallines
      \draw (UL) -- ++(180:0.65) node[anchor=east] {$r{+}1$};
      \draw[dotted] (UL)+(170:0.5) to [bend left=45] ++(100:0.5);
      \draw (UL) -- ++(90:0.65) node[anchor=south] {$s{-}1$};
      \draw[double] (UR) -- ++(45:0.65);
      \draw (LR) -- ++(0:0.65) node[anchor=west] {$s$};
      \draw[dotted] (LR)+(-10:0.5) to [bend left=45] ++(-80:0.5);
      \draw (LR) -- ++(-90:0.65) node[anchor=north] {$r{-}1$};
      \draw (LL) -- ++(-135:0.65) node[anchor=north east] {$r$};
    \end{tikzpicture}}}
    &=
    \frac{
        \langle r\,(s{-}1)^-\,s^-\,s\rangle
        \langle r\,s{-}1\,s\,(s-1)^-\rangle}
        {\langle r^+\,s{-}1\,s\,r\rangle
        \langle s\,s^-\,s{-}1\,(s{-}1)^-\rangle}
    [(s{-}1)^-,s^-,s{-}1,s,r]\, .
    \end{aligned}
    \label{eq:supp-Rrss-general}
\end{equation}
In our notation, this boundary invariant is denoted simply by
$R_{rss}\equiv R'_{rss}$.

The two prefactors used in the main text are obtained from
Eqs.~\eqref{eq:supp-Rrst-general} and \eqref{eq:supp-Rrss-general}.  The
non-boundary representative is
\begin{equation}
	R_{241}= \raisebox{-3em}{\begin{tikzpicture}[thick, scale=0.74]
		\drawLLblack
		\drawULwhite
		\drawURblack
		\drawLR{$\scriptstyle{0}$}
		\drawboxinternallines
		\draw (UL) -- ++(135:0.65) node[above left=-2pt] {$2$};
		\draw (UR) -- ++(45:0.65) node[above right=-2pt] {$3$};
		\draw (LR) -- ++(0:0.65) node[right=-2pt] {$4$};
		\draw[double] (LR) -- ++(-90:0.65);
		\draw (LL) -- ++(-135:0.65) node[below left=-2pt] {$1$};
	\end{tikzpicture}} = [2,3,4,4^-,1] \, .
\end{equation}
The boundary representative is 
\begin{equation}
	 R_{144} = \raisebox{-3em}{\begin{tikzpicture}[thick, scale=0.74]
		\drawLLblack
		\drawULwhite
		\drawUR{$\scriptstyle{0}$}
		\coordinate (LR) at ( \boxsize, -\boxsize);
		\drawboxinternallines
		\draw (UL) -- ++(135:0.65) node[above left=-2pt] {$1$};
		\draw (UR) -- ++(0:0.65) node[right=-2pt] {$3$};
		\draw (UR) -- ++(90:0.65) node[above=-2pt] {$2$};
		\draw[double] (LR) -- ++(-45:0.65);
		\draw (LL) -- ++(-135:0.65) node[below left=-2pt] {$4$};
	\end{tikzpicture}}
=\frac{
		\langle 1\,3^-\,4^-\,4\rangle
		\langle 1\,3\,4\,3^-\rangle}{\langle 1^+\,3\,4\,1\rangle
		\langle 4\,4^-\,3\,3^-\rangle} [3^-,4^-,3,4,1] = 
	\raisebox{-3em}{\begin{tikzpicture}[thick, scale=0.74]
		\drawLLwhite
		\drawULblack
		\drawURwhite
		\drawLR{$\scriptstyle{1}$}
		\drawboxinternallines
		\draw (UL) -- ++(135:0.65) node[above left=-2pt] {$2$};
		\draw (UR) -- ++(45:0.65) node[above right=-2pt] {$3$};
		\draw (LR) -- ++(0:0.65) node[right=-2pt] {$4$};
		\draw[double] (LR) -- ++(-90:0.65);
		\draw (LL) -- ++(-135:0.65) node[below left=-2pt] {$1$};
	\end{tikzpicture}} \, ,
\end{equation}
Thus, $R_{144}$ is the parity conjugate of $R_{241}$; this relation becomes
manifest after replacing $R_{144}$ by the equivalent on-shell diagram in the
last equality~\cite{Bianchi:2018rrj}.

The remaining prefactor in \eqref{eq:cycansatz}, denoted as $S_1=R_{12,34}$ in the
main text, is the algebraic leading singularity
associated with the finite three-mass triangle in the four-point NMHV one-loop
form factor.  Following Ref.~\cite{Bianchi:2018rrj}, 
\begin{equation}
	R_{12,34}
	=\raisebox{-3.5em}{\begin{tikzpicture}[thick, scale=0.72]
        \coordinate (T1) at ( 0.45 ,  0.78);
        \coordinate (T2) at ( 0.45 , -0.78);
        \coordinate (T3) at (-0.9 ,  0.0 );
        \draw (T1) -- (T2) -- (T3) -- (T1);
        \draw (T1) -- ++( 60:0.6) node[above right=-2pt] {$1$};
        \draw (T1) -- ++(  0:0.6) node[right] {$2$};
        \draw (T2) -- ++(  0:0.6) node[right] {$3$};
        \draw (T2) -- ++(-60:0.6) node[below right=-2pt] {$4$};
        \draw[double] (T3) -- ++(180:0.6);
        \node[circle, fill=black!20, draw, minimum size=0.42cm, inner sep=1pt] at ( 0.45 ,  0.78) {$\scriptstyle{0}$};
        \node[circle, fill=black!20, draw, minimum size=0.42cm, inner sep=1pt] at ( 0.45 , -0.78) {$\scriptstyle{0}$};
        \end{tikzpicture}}=
	\frac{1}{2}
	\left(
	\mathcal{L}_{1,3}(\ell^{*})
	+\mathcal{L}_{1,3}(\bar \ell^{*})
	\right)
	\frac{\sqrt{u_1 u_3}}{r_2}\, ,
	\label{eq:supp-R1234}
\end{equation}
where
\begin{equation}
	\begin{aligned}
		\mathcal{L}_{r,s}(\ell)
		={}[\ell,r,r{-}1,r^-,(r{-}1)^-]\,\times \,
		\frac{
			\langle \ell\,r\,r{-}1\,r^-\rangle
			\langle \ell\,r^-\,(r{-}1)^-\,r{-}1\rangle}
		{\langle \ell\,r\,r{-}1\,s{-}1\rangle
			\langle \ell\,r^-\,(r{-}1)^-\,s\rangle}\times\frac{
			\langle s{-}1\,s\,r{-}1\,r\rangle^{1/2}
			\langle s{-}1\,s\,(r{-}1)^-\,r^-\rangle^{1/2}}
		{\langle r{-}1\,r\,(r{-}1)^-\,r^-\rangle}\, ,
	\end{aligned}
	\label{eq:supp-Rrs}
\end{equation}
and $\ell^{*}$ and $\bar{\ell}^{*}$ are the two solutions of the following cut equations:
\begin{equation}
	\langle \ell \, 4^- 1 \rangle = \langle \ell \,  23 \rangle = \langle \ell \,  4 1^+ \rangle = \langle \ell \, I_{\infty} \rangle = 0.
\end{equation}
The individual solutions contain square-root dependence.  In the symmetric
combination in Eq.~\eqref{eq:supp-R1234}, this dependence reduces to a single
overall square root, carried by $r_2$, in $R_{12,34}$.

\section{Symbol Alphabet of 88 Letters}
The four-point MHV form factor has recently been bootstrapped through four loops~\cite{MHV:fourloop}. In the notation of the ancillary file \texttt{aAlphabet.m} of \cite{Dixon:2024yvq}, which contains a 93-letter alphabet, all available MHV results through four loops involve only an 88-letter subset. This suggests that the 88-letter subset is a stable alphabet for the four-point form factor. We have also rewritten our NMHV result in the same notation.

This subset is obtained from the 93-letter list by removing five letters: $\{a_{43},a_{44},a_{73},a_{76},a_{87}\}$. Here we list this 88-letter subset using cyclically independent seeds; all cyclic images are understood. It consists of 68 rational letters and 20 algebraic letters. The rational letters can be written in terms of four-brackets of periodic momentum twistors. The cyclically independent rational seeds are
\begin{equation}
    \begin{aligned}
        &\langle 1234\rangle,\,
        \langle 1241^+\rangle,\,
        \langle 121^+2^+\rangle,\,
        \langle 1231^+\rangle,\,
        \langle 1232^+\rangle,\,
        \langle 1233^+\rangle,\,
        \langle 1242^+\rangle,\,
        \langle 1343^+\rangle,\,
        \langle 131^+3^+\rangle,
        \\
        &\langle (1^+2^+)\cap(123)\,3^+\,4^{+}\,1^{++}\rangle,\,
        \langle 3^+4^+\,(234)\cap(41^+2^+)\rangle,
        \\
        &\langle (41^+)\cap(2^+3^+4^+)\,2\,
        (4^+1^{++})\cap(2^{++}3^{++}4^{++})\,2^+\rangle,
        \\
        &\langle 1\,(23)\cap(41^+2^+)\,1^+\,
        (2^+3^+)\cap(4^+1^{++}2^{++})\rangle,\,
        \langle (41^+)\cap(123)\,2\,
        (4^+1^{++})\cap(1^+2^+3^+)\,2^+\rangle,
        \\
        &\langle (12)\cap(1^+2^+3^+)\,3\,4\,2^+\rangle,\,
        \langle (23)\cap(41^+2^+)\,2^+\,3^+\,1^{++}\rangle,\,
        \langle (41^+)\cap(123)\,2\,2^+\,3^+\rangle,\,
        \langle (41^+)\cap(123)\,2^+\,4^+\,1^{++}\rangle .
    \end{aligned}
    \label{eq:supp-A88-rat}
\end{equation}
Some of these rational letters contain intersections of lines and planes in twistor space. Here $(ab)$ denotes the line through $Z_a$ and $Z_b$, while $(abc)$ denotes the plane through $Z_a,Z_b,Z_c$. Then $\cap$ denotes the projective intersection: $(ab)\cap(cde)$ is the twistor where the line $(ab)$ meets the plane $(cde)$, and $(abc)\cap(def)$ is the line common to two
planes. These are composite twistors built from the external kinematics. For example,
\begin{equation}
    (ab)\cap(cde)
    =
    Z_a\langle b c d e\rangle
    -
    Z_b\langle a c d e\rangle ,
    \label{eq:supp-cap-point}
\end{equation}
and the corresponding four-brackets can be expanded as
\begin{equation}
    \begin{aligned}
    \langle (ab)\cap(cde)\, f\,g\,h\rangle
    &=
    \langle a f g h\rangle \langle b c d e\rangle
    -
    \langle b f g h\rangle \langle a c d e\rangle ,
    \\
    \langle a\,b\,(cde)\cap(fgh)\rangle
    &=
    \langle a c d e\rangle \langle b f g h\rangle
    -
    \langle b c d e\rangle \langle a f g h\rangle .
    \end{aligned}
    \label{eq:supp-cap-bracket}
\end{equation}
The algebraic letters involve two cyclically independent square roots, denoted by $r_1$ and $\Sigma_1$. For each root there are five algebraic seed letters; after including cyclic images these give the remaining 20 letters,
\begin{equation}
    \begin{aligned}
        r_1 : \quad & \frac{v_1+v_2+r_1}{v_1+v_2-r_1}, \, \frac{2-v_1-v_2+r_1}{2-v_1-v_2-r_1}, \, \frac{v_1-v_2+r_1}{v_1-v_2-r_1}, \, \frac{v_1-v_2-2u_1+2u_3+r_1}{v_1-v_2-2u_1+2u_3-r_1}, \frac{P_1 + r_1 \text{tr}_5}{P_1 - r_1 \text{tr}_5}
        \\
        \Sigma_1 : \quad & \frac{u_3 v_1-u_1 v_2+u_1+u_1 u_2-u_2 u_3-v_1+v_1 v_2 + \Sigma_1}{u_3 v_1-u_1 v_2+u_1+u_1 u_2-u_2 u_3-v_1+v_1 v_2-\Sigma_1},\\
        & \frac{-2 u_1 v_1-2 u_2 v_1+u_3 v_1-u_1 v_2+u_1+u_1 u_2-u_2 u_3+2 v_1^2+v_2 v_1-v_1+\Sigma_1}{-2 u_1 v_1-2 u_2 v_1+u_3 v_1-u_1 v_2+u_1+u_1 u_2-u_2 u_3+2 v_1^2+v_2 v_1-v_1-\Sigma_1},\\
        & \frac{-u_3 v_1-u_1 v_2-2 u_2 v_2+u_1-u_1 u_2+2 u_2+u_2 u_3-v_1+v_1 v_2+\Sigma_1}{-u_3 v_1-u_1 v_2-2 u_2 v_2+u_1-u_1 u_2+2 u_2+u_2 u_3-v_1+v_1 v_2-\Sigma_1},\\
        & \frac{2 u_1 v_1-u_1 v_2-u_3 v_1-2 u_1^2-u_2 u_1+2 u_3 u_1+u_1+u_2 u_3-v_1+v_1 v_2+\Sigma_1}{2 u_1 v_1-u_1 v_2-u_3 v_1-2 u_1^2-u_2 u_1+2 u_3 u_1+u_1+u_2 u_3-v_1+v_1 v_2-\Sigma_1}, \frac{P_2 + \Sigma_1 \text{tr}_5}{P_2 - \Sigma_1 \text{tr}_5},
    \end{aligned}
    \label{eq:supp-A88-alg}
\end{equation}
where 
\begin{equation}
\begin{aligned}
P_1 &=-u_1 u_2 v_1+u_3 u_2 v_1-u_2 v_1+u_1 u_2 v_2-u_3 u_2 v_2-2 u_2 v_1 v_2-u_2 v_2-u_3 v_1^2\\&-u_1 v_2^2-u_1 v_1 v_2-u_3 v_1 v_2+2 u_2^2+2 u_1 u_2+2 u_3 u_2+v_1 v_2^2+v_1^2 v_2,
\end{aligned}
\end{equation}
and 
\begin{equation}
\begin{aligned}
P_2 &= -u_1^2 v_2^2+2 u_2 u_1^2 v_2+u_1^2 v_2+2 u_1 v_1 v_2^2+u_2 u_1 v_1-2 u_2 u_3 u_1 v_1\\&+u_3 u_1 v_1-u_2 u_1 v_2-2 u_2 u_3 u_1 v_2-2 u_2 u_1 v_1 v_2-2 u_3 u_1 v_1 v_2-2 u_1 v_1 v_2-u_3^2 v_1^2\\&-u_3 v_1^2+2 u_2 u_3^2 v_1-u_2 v_1+2 u_3 v_1^2 v_2+u_2 v_1 v_2-2 u_2 u_3 v_1 v_2-u_2^2 u_1^2\\&-u_2 u_1^2+u_2^2 u_1+u_2 u_1+2 u_2^2 u_3 u_1+3 u_2 u_3 u_1-u_2^2 u_3^2+u_2^2 u_3-v_1^2 v_2^2+v_1^2 v_2.
\end{aligned}
\end{equation}

The two-loop NMHV result contains only 78 of these 88 letters and includes the two-loop MHV alphabet as a subset. The letters absent from the two-loop result are $\{a_{27},a_{28},\ldots,a_{34},a_{55},a_{56}\}$. Here $a_{55} = \Sigma_1^2$ and $a_{56} = \Sigma_2^2$ are the two ``square-root" letters mentioned in the main text.

\section{Details of the Three-Loop Bootstrap}

In the three-loop bootstrap we work with the minimally-subtracted hard
function.  We use ``hard function'' in the standard infrared-factorization
sense: the loop-level form factor is written as
\begin{equation}
    \mathcal F_{4,k}=Z_4\,H_{4,k},
\end{equation}
where \(Z_4\) is the universal infrared subtraction factor in the
minimal-subtraction scheme and is independent of the Grassmann degree \(k\).
Thus \(H_{4,k}=Z_4^{-1}\mathcal F_{4,k}\) is finite.  Since the same \(Z_4\)
multiplies the MHV and NMHV sectors, the NMHV hard function is related to the
ratio function by
\begin{equation}
    H_{4,1}=H_{4,0}\mathcal R_4 .
\end{equation}
In particular, at three loops,
\begin{equation}
\begin{aligned}
H_{4,1}^{(3)}
={}&\mathcal R_4^{(3)}
+H_{4,0}^{(1)}\mathcal R_4^{(2)}
+H_{4,0}^{(2)}\mathcal R_4^{(1)}
+H_{4,0}^{(3)}\mathcal R_4^{(0)} ,
\end{aligned}
\end{equation}
with the conventional tree-level normalization \(H_{4,0}^{(0)}=1\).

We impose extended Steinmann conditions on the hard-function symbol.  In both
the two-loop NMHV and three-loop MHV four-point form factors, adjacent pairs
\(v_i\otimes v_j\) with \(i\neq j\) are forbidden in the first two entries.
For later entries, a weaker extended Steinmann condition allowing pairs of the
form \(v_i\otimes v_{i+2}\) is satisfied, while the last two entries satisfy
the full Steinmann restriction.  When constructing the integrable symbols, we
impose the strict Steinmann relation on the first two entries, but only the
weaker extended Steinmann conditions on the later entries.  The strict
Steinmann relation is still satisfied for the last two entries of the
three-loop result.

The weight-six space built from the 88-letter alphabet, integrability, and the
extended Steinmann adjacency condition contains 54,603 symbols before
multiplication by the three classes of NMHV leading singularities.  After
imposing Galois, parity, and dihedral symmetry, the number of unknowns is
reduced to \(3817\,(f_s)+17335\,(f_+)+1089\,(f_-)=22241\).  The remaining
constraints include spurious-pole cancellation and the multi-collinear limits. When imposing the multi-collinear limits, we
first require them to be well defined, meaning that all symbol letters absent
from the corresponding target function cancel after the limit is taken.  These
constraints then fix the hard function symbol uniquely.  We then check the soft
and double-soft limits. Finally, after converting the result back to the ratio function, we verify DDCI by
checking that the \(q^2\to0\) limit is finite and depends only on the ten
directional-dual-conformal letters \cite{Chicherin:2018wes}.

\section{Ancillary Files}

We provide the symbol-level results for the two- and three-loop NMHV ratio
function and hard function in the ancillary files.  All data are stored in
Wolfram Expression Format (WXF).  The symbol letters are written in the notation of the
93-letter \(a\)-alphabet of Ref.~\cite{Dixon:2024yvq}.  More precisely, all
entries are restricted to the 88-letter subset discussed above: an integer
\(i=1,\ldots,88\) in the data denotes the \(i\)-th letter in the ordered list
obtained from \(a_1,\ldots,a_{93}\) after removing
\(a_{43},a_{44},a_{73},a_{76},a_{87}\).

\begin{table}[h]
\centering
\small
\setlength{\tabcolsep}{10pt}
\renewcommand{\arraystretch}{1.18}
\begin{tabular}{ll}
\hline\hline
File(s) & Content \\
\hline
\texttt{NMHVFF4pt2LRatioFunction.wxf} & two-loop ratio function \\
\texttt{NMHVFF4pt2LHardFunction.wxf} & two-loop hard function \\
\texttt{NMHVFF3L4pt\_<type>\_cop1234.wxf} & first four entries at three loops \\
\texttt{NMHVFF3L4pt\_<type>\_cop56\_<part>.wxf} & last two entries at three loops \\
\hline\hline
\end{tabular}
\end{table}

The two-loop data are given in
\texttt{NMHVFF4pt2LRatioFunction.wxf} and
\texttt{NMHVFF4pt2LHardFunction.wxf}, for the ratio function and hard function,
respectively.  Each file contains a five-element Mathematica \texttt{List}, ordered as
\begin{equation}
    \{ f_+, f_{\mathrm{odd},1}, f_{\mathrm{odd},2}, f_{\mathrm{odd},3}, f_s \}.
\end{equation}
Here \(f_+\) is the parity-even function,
\(f_{\mathrm{odd},1}\), \(f_{\mathrm{odd},2}\), and
\(f_{\mathrm{odd},3}\) are the three independent parity-odd combinations, and
\(f_s\) is the function multiplying the algebraic leading singularity.  This
ordering follows the parity organization in the main text.  Each
element of the \texttt{List} gives the corresponding weight-four symbol as a
\texttt{SparseArray}.

At three loops, the weight-six symbols are stored in coproduct form.  For each
of the ratio function and hard function, the file
\texttt{NMHVFF3L4pt\_<type>\_cop1234.wxf} contains the first four
entries of the symbol, where \texttt{<type>} is either \texttt{Hard} or
\texttt{Ratio}.  The files
\texttt{NMHVFF3L4pt\_<type>\_cop56\_<part>.wxf}, with
\(\texttt{<part>}=\texttt{even},\texttt{odd1},\texttt{odd2},\texttt{odd3},\texttt{s}\),
contain the last two entries for the five independent components.  The full
weight-six symbol for each component is obtained by multiplying the
corresponding first-four-entry block and last-two-entry block.  For example, in
Mathematica the parity-even component of the three-loop hard function can be
reconstructed as
\begin{verbatim}
hardCop1234   = Import["NMHVFF3L4pt_Hard_cop1234.wxf"];
hardCop56Even = Import["NMHVFF3L4pt_Hard_cop56_even.wxf"];
hardEven3L    = SparseArray[hardCop1234 . hardCop56Even];
\end{verbatim}
The dot product contracts the common coproduct index and reconstructs the full
\texttt{SparseArray} representation of the weight-six symbol.  The
ratio-function data and the other components are reconstructed analogously.